\documentclass[aps,nofootinbib,twocolumn,pra,10pt]{revtex4-2}

\usepackage{amsmath}
\usepackage{amssymb}
\usepackage{braket}
\usepackage{graphicx}
\usepackage{ulem}
\usepackage{xcolor}
\usepackage{xspace}

\newcommand{\PMD}{\mathcal{D}}
\newcommand{\rmi}{\mathrm{i}}
\newcommand{\rmd}{\mathrm{d}}

\definecolor{darkgreen}{rgb}{0,.7,0}

\newcommand{\asec}{asec\xspace}

\newcommand{\IL}{\ensuremath{I_{\mbox{\sc
l}}}\xspace}
\newcommand{\xin}{\ensuremath{{x_{\mbox{\tiny in}}}}\xspace}
\newcommand{\xion}{\ensuremath{{x_{0}}}\xspace}

\newcommand{\tion}{\ensuremath{t_0}\xspace}

\newcommand{\Ip}{\ensuremath{{I_p}}\xspace}

\newcommand{\tauB}{\ensuremath{\tau}\xspace} 
\newcommand{\taup}{\ensuremath{\tau^\star}\xspace} % 
\newcommand{\taupol}{\ensuremath{\tilde\tau}\xspace} % 

\newcommand{\LCPMR}{Sorbonne Universit\'{e}, CNRS, Laboratoire de Chimie Physique – Mati\`{e}re et Rayonnement, LCPMR, 75005 Paris, France}

\begin{document}
\title{Attosecond tunneling time measurements through momentum squeezing in strong field ionization}

\author{Jonathan Dubois} 
\author{Léonardo Rico} 
\author{Camille L\'{e}v\^{e}que} 
\author{J\'{e}r\'{e}mie Caillat}
\author{Richard Ta\"{i}eb} \affiliation{\LCPMR}

\maketitle

{\bf 
Tunneling of a particle through a potential barrier is a fundamental physical process and a major thought-provoking outcome of quantum physics. It is at the basis of multiple scientific and technological advances and strongly influences both the structuring and the dynamics of matter at the microscopic scale~\cite{Binning1987, Huillier1993, Trixler2013, Wild2023}. Without a classical counterpart, it defies our intuitive perception and understanding of the motion of a particle. Thus, the temporal  characterization of tunneling, typically in terms of the time spent `under the barrier', referred to as \textit{tunneling time}, raises several debates and questions on its interpretation and measurability~\cite{Buttiker1982, Sokolovski1995, Yamada2004, Ramos2020, Sokolovski1997, Sokolovsky2018, Sainadh2019, Sainadh2020, Hofmann2021}.
Here we show that an electron wavepacket tunneling out of an atom through the potential barrier induced by a strong electric field, carries in its momentum profile the value of the corresponding tunneling time, in a self-probing manner. 
In a revisited interpretation of the attoclock setup~\cite{Eckle2008, Torlina2015, Sainadh2019, Dianxiang2023}, we view a circularly polarized light pulse as a temporal prism which maps the barrier configuration, and hence the tunneling dynamics, onto different photoelectron ejection directions. From our simulations, we find that tunneling times in the infrared regime are of the order of hundreds of attoseconds, in agreement with previous theories ~\cite{Buttiker1982, Sokolovski1995, Yamada2004, Ramos2020, Sokolovski1997, Sokolovsky2018}.
}

\begin{figure}[t]
    \centering
    \includegraphics[width=.5\textwidth]{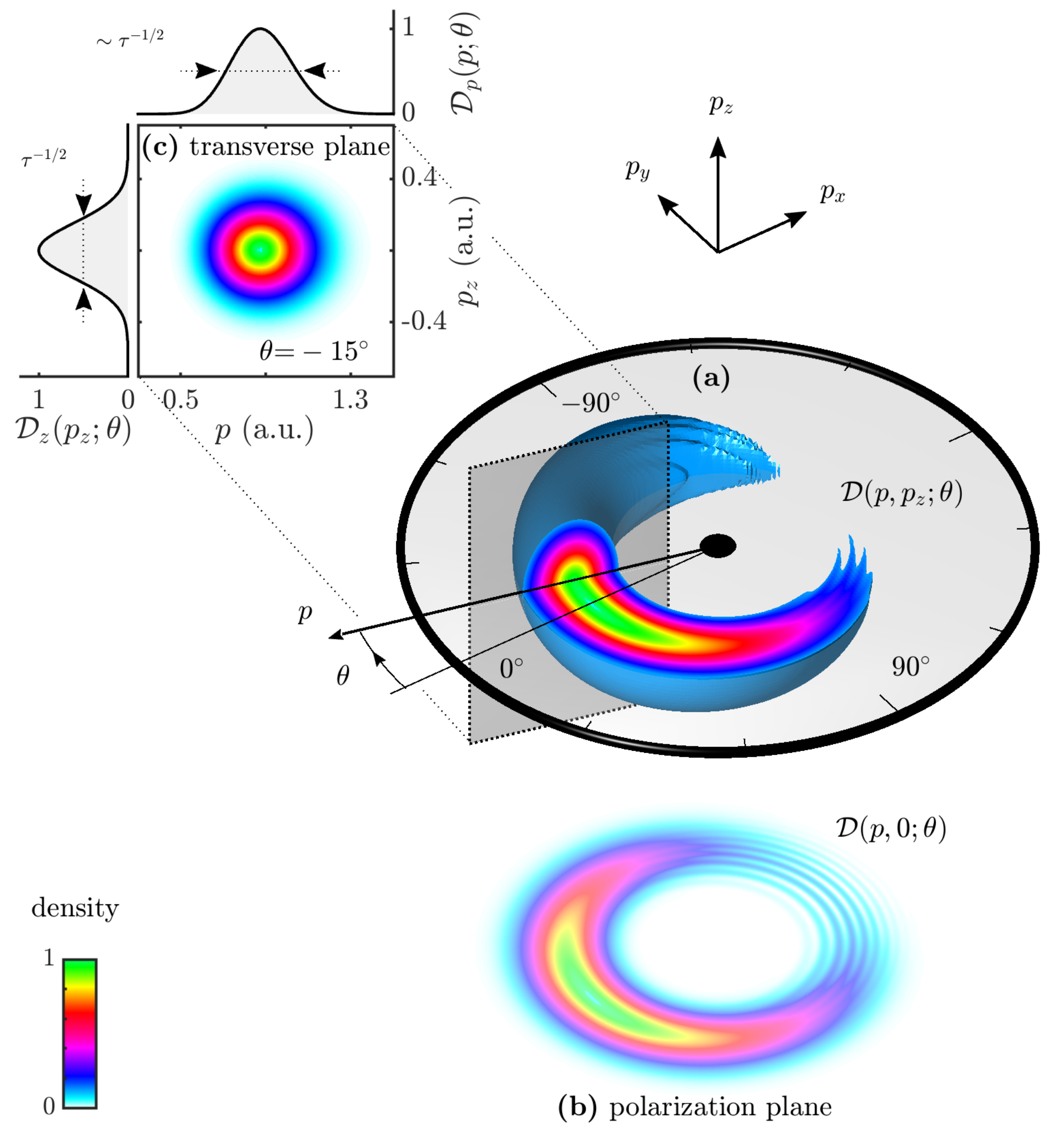}
    \caption{Illustrative photoelectron momentum distributions resulting from strong field ionization of H with a circularly polarized short laser pulse ($\lambda{=}800$ nm,  $\IL{=}10^{14} \, \rm W \ cm^{-2}$) simulated using SFA. The PMDs are represented as functions of the final photoelectron momentum expressed in cylindrical coordinates, i.e., $\mathbf{p}{=}p (\mathbf{e}_x \cos \theta{+}\mathbf{e}_y \sin \theta) {+}\mathbf{e}_z p_z$ where $p_z$ is the component along the light propagation direction $\mathbf{e}_z$, $p$ is the modulus in the transverse plane $(\mathbf{e}_x,\mathbf{e}_y)$ and the polar angle $\theta$ represents the photoemission direction. (a) Full distribution. The black line indicates the reference angle $\theta{=}0^\circ$, indicating the orientation of $-\mathbf{A}(t)$ when the pulse envelop reaches its maximum. (b) Cut of the PMD in the polarization plane. 
    (c) Cut and projections of the PMD in the transverse plane at an illustrative angle  $\theta{=}-15^{\circ}$. The widths of the distributions in the transverse plane encode the duration \tauB of the tunneling process experienced by the photoelectrons detected in each direction $\theta$.} 
    \label{fig:illustration}
\end{figure}
The puzzling temporal aspects of quantum tunneling  have been investigated both theoretically and experimentally in atomic and optical physics. Formally, various  `tunneling time' formulas have been derived and investigated, revisiting standard tools of wave physics~\cite{Yamada2004,Sokolovski1995,Sokolovski1997} or semi-classical concepts~\cite{Buttiker1982,Buttiker1983}. They are ideally addressed in terms of a tunneling {\it duration} (hereafter denoted \tauB) and of an {\it instant} of tunnel exit (\tion), the latter implicitly referring to a case-specific external clock. Using a variety of experimental setups, studies revealed a broad range of time scales. For example,  in so-called Larmor clocks~\cite{Martin1992, Ramos2020, Spierings2021}, the spin precession induced by a  magnetic field  in the tunneling region serves as an auxiliary clock revealing durations of few ms~\cite{Ramos2020}, while, in cold atom lattice systems~\cite{Fortun2016, Ramos2020}, $\mu$s durations were measured as delays between tunneled and reflected wavepackets. Measurements based on the transversal displacements of light  beams transmitted through an air slab embedded in glass revealed tunneling durations on the fs scale~\cite{Balcou1997}.

With the advent of intense laser sources~\cite{Strickland1985}, strong field atomic and molecular physics emerged as a singular playground to investigate {\it in situ} the tunneling of electrons in matter at the attosecond timescale~\cite{krause1992,Corkum1993,Lewenstein1994}. In this framework, elaborate experimental schemes were proposed to extract tunneling timings from observable quantities such as the spectrum and phase of high order harmonics generated with bichromatic pulses~\cite{Pedatzur2015}, or the photoelectron spectra resulting from strong field ionization of atoms with circularly polarized field~\cite{Eckart2018,Eckart2018b}. The present work focuses on the measurement of tunneling times out of photoelectron momentum distributions in this context.

When an atom is subjected to a strong infrared (IR) laser pulse, one of its valence electrons can be pulled out by tunneling through the transient effective barrier created by the field. Ground-breaking progress towards the measurement of tunneling times has been achieved over the last two decades, using the so-called Attoclock technique~\cite{Eckle2008, Torlina2015, Ni2016, Sainadh2019, Hofmann2021, Dianxiang2023}. 
In this experimental scheme, angular streaking with short circularly or elliptically polarized pulses are used to encode the photoemission dynamics in the photoelectron angular distribution. The reported times range up to few tens of attoseconds. As noted in the pioneering work by Eckle {\it et al}~\cite{Eckle2008}, this is one order of magnitude below the expected values. This initiated a series of  debates and controversies fueled over the years by improved experiments and refined interpretations. Assigning a singular role to the Coulomb potential notably decreased the measured times to near-zero~\cite{Torlina2015, Ni2016, Sainadh2019,Sainadh2020,Hofmann2021}. Based on the \textit{hypothesis} that the photoelectron most probably enters the potential barrier at the maximum of the electric field, it was thus concluded in~\cite{Eckle2008, Torlina2015, Sainadh2019} that tunneling is instantaneous. This remained in stunning conflict with the results reported in~\cite{Balcou1997, Fortun2016, Ramos2020} and highlights ambiguities in the assumptions on how \tion and \tauB relate to each other and to the observables. Additional simulations using virtual detectors~\cite{Teeny2016, Wang2018} indicated that the photoelectron wavepacket most probably {\it exits} the potential barrier near the time of maximum  electric field, and that it {\it enters} the barrier few hundred attoseconds earlier.

In light of these contradicting results, we provide a rigorous theoretical framework that reconciles the previous findings and propose a protocol to unambiguously extract tunneling durations and exit times from Attoclock experiments. 
We establish that the complete tunneling dynamics associated with each photoemission direction, in a circularly polarized strong field, is directly encoded in the shape of the photoelectron momentum distribution (PMD). This manifests in a transparent way when modeling the process in cylindrical coordinates, where the PMD factorizes as $\PMD (p,p_z;\theta){=}  \PMD_z (p_z;\theta){\times}\PMD_p (p;\theta)$. Here, $p_z$ is the momentum coordinate along the light propagation direction, i.e., perpendicular to the potential vector $\textbf{A}(t)$ at all times, and $p=(p_x^2+p_y^2)^{1/2}$, see Fig.~\ref{fig:illustration}.

Using the strong field approximation (SFA)~\cite{Lewenstein1994}, we found that the most probable duration \tauB associated with each photoemission angle $\theta$ appears redundantly in the distribution projection as
\begin{subequations}
\begin{eqnarray}
\label{eq:distribution_z}
    \PMD_z (p_z ; \theta ) &\propto& e^{-\frac{1}{\hbar m} \tauB(\theta) \: p_z^2} \\
\label{eq:distribution_p}
    \PMD_p (p ; \theta ) &\propto& e^{-\frac{1}{\hbar m}\taupol(\theta) \: [ p-p_0 (\theta) ]^2}
\end{eqnarray}
%(Il me semble que c'est $1/\hbar m$ si on parle de quantité de mouvement et $\hbar/m$ si on parle de vecteur d'onde $\mathbf{k}$).}
where $\hbar$ is the reduced Planck's constant and $m$ is the electron mass. 
In these expressions, $\tauB (\theta)$ corresponds, in each direction $\theta$, to the tunneling duration associated with the most probable momentum $(p,p_z) = (p_0 , 0)$.
The extraction of \tauB out of the PMD profile along $p_z$ is straightforward, while it is slightly more elaborate along $p$, through the inverse width 
\begin{eqnarray}
    \label{eq:width_dt}
    \taupol (\theta) = \tauB \times \left(
\frac{\omega\tauB\cosh(2\omega \tauB)-\sinh(2\omega\tauB)}{\omega \tauB\sinh^2 (\omega \tauB)}+\frac{1}{(\omega\tauB)^2}
   \right)
\end{eqnarray}
where $\omega$ is the laser frequency and the $\theta$-dependency of \tauB is implicit.
\end{subequations}
As for the exit time \tion, it is directly given by the photoemission angle $\theta$ -- opposite to the orientation of the potential vector $\mathbf{A}(\tion)$ at the instant of tunnel exit -- through the simple one-to-one correspondence 
\begin{equation}\label{eq:t0}
\theta=\omega\tion
\end{equation}
which is essential to the Attoclock scheme, see~\cite{Torlina2015} and references therein.

The detailed derivations of  Eqs.~\eqref{eq:distribution_z}-\eqref{eq:width_dt} are provided in the Supplementary Material (SM). In few words, we built our reasoning by explicitly writing the time parameter in the compact form
\begin{equation}
\label{eq:tstar}
t^\star=\tion+i\tauB
\end{equation}
which provides a comprehensive temporal characterization of the tunnel process~\cite{PerelomovI1966,PerelomovII1967,PerelomovIII1967,Pedatzur2015}.
Furthermore, our derivations results in the 
transcendental equation~\cite{PerelomovII1967, Barth2011}
\begin{equation}
\label{eq:gamma_phi}
     - \left( \dfrac{\sinh ( \omega \tauB )}{\omega \tauB} \right)^2 + \dfrac{\sinh (2 \omega \tauB)}{\omega \tauB}  - 1 = \gamma (\tion)^2 ,
\end{equation}
which associates a unique \tauB to a given \tion, where $\gamma (\tion) {=} \omega \sqrt{2 m I_p} / F_0 f(\tion)$ is the instantaneous Keldysh parameter, $I_p$ is the ionization potential, $F_0$ is the electric field strength and $f(t)$ its temporal envelope. 

Remarkably, we alternatively obtained Eqs~\eqref{eq:distribution_z}-\eqref{eq:width_dt} in the Wentzel–Kramers–Brillouin framework~\cite{Messiah1961} by  identifying \tauB as the `traversal time' established by B\"uttiker and Landauer~\cite{Buttiker1982}. 
For an electron tunneling through a potential barrier $V(x)$ with energy $E$, it reads
\begin{equation}
\label{eq:Buttiker}
    \tauB = \int\limits_\xin^\xion \dfrac{\rmd x}{\sqrt{2 [ V(x) - E ]/m}}
\end{equation}
where \xin and \xion are the tunnel entrance and exit coordinates, between which $E<V(x)$. This approach (see SM) confers to the measurable \tauB  an intuitive insight highlighted e.g. in~\cite{Balcou1997}.

\begin{figure*}[t]
    \centering
    \includegraphics[width=.9\textwidth]{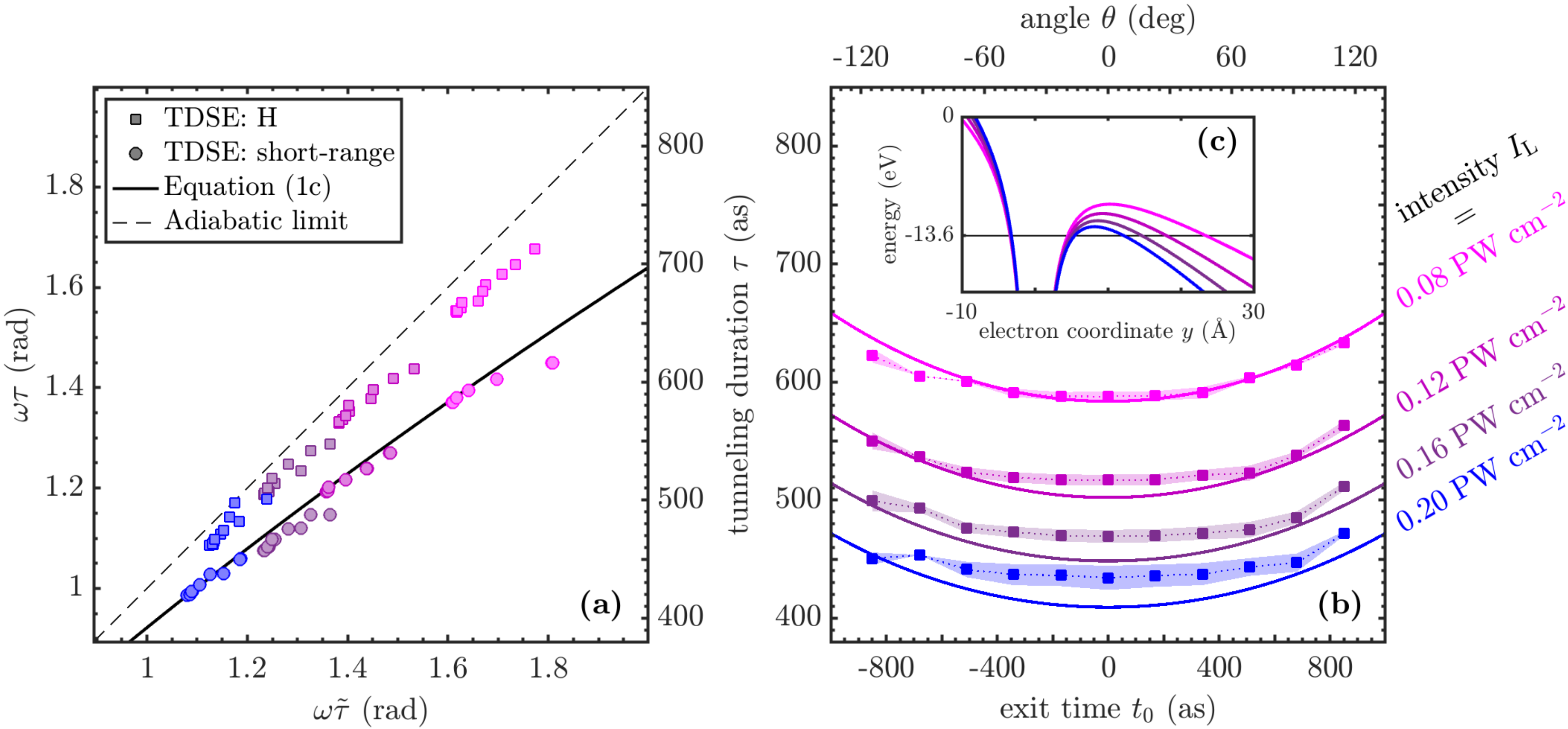}
    \caption{Tunneling times in simulated strong field ionization. (a) Correlation between the inverse widths of the PMD cuts along the $p_z$ and $p$ coordinates,  respectively $\omega\tauB$ and $\omega\tilde\tauB$. The shaded squares indicate the data extracted from TDSE simulations on H with a 800-nm laser, at 11 photoemission angles, with 4 different laser intensities [see the $\theta$-sampling and the \IL-dependent color code in frame (b)]. The circles show the equivalent data obtained with a model atom built on a short range potential. The solid line represents the theoretical correlation given by Eq.~\eqref{eq:width_dt}. The dashed line indicates the diagonal ($\omega\tauB=\omega\tilde\tauB$), which physically corresponds to the expected correlation in the adiabatic limit.
    (b) Tunneling duration as a function of the photoemission angle $\theta$ and the corresponding tunnel exit time $t_0{=}\theta/\omega$. 
    The squares indicate the simulated data in H obtained by fitting Eq.~\eqref{eq:width_dt} to the PMDs. The shaded areas represent the statistical errors on the fitting parameters. The solid lines represent the theoretical values given by Eq.~\eqref{eq:gamma_phi}. The inset (c) shows the shape of the  effective potential barrier along the laser electric field direction for each intensities [same color code as in frame (b)] at the peak amplitude of the laser field ($\theta{=}0$).}
\label{fig:tunneling_times}
\end{figure*}

The parametrization of the transverse PMD cuts given by Eqs.~\eqref{eq:distribution_z}-\eqref{eq:width_dt} puts forward a linear shrinking of the momentum distributions as \tauB increases. This can be intuitively interpreted in the semi-classical adiabatic picture. For a given barrier configuration and a given $\theta$, photoelectrons departing from the shortest tunneling path, perpendicular to $\theta$, spend more time under the barrier and exit with a decreased probability. Due to the exponential nature of the tunnel decay, the longer the duration of the main path (\tauB), the more pronounced the transverse attenuation and the more peaked the exit distribution. 

The less straightforward $\tau$-dependency of the PMD width along $p$ [Eq.~\eqref{eq:distribution_p}] also boils down to a bijective one-to-one map. Its  elaborate structure results from the correlation of the momentum coordinates in the polarization plane, where the $\mathbf{A}(t)$ evolves {\it during tunneling}. In contrast, the motion in the transverse $p_z$ direction is free within the SFA.  The factor between parentheses on the r.h.s. of Eq.~\eqref{eq:width_dt} reaches 1 for small $\omega \tauB$, and therefore $\taupol {\approx} \tauB$ in this limit. 
Hence, an elliptical distortion of the shape of $\PMD (p, p_z ; \theta )$ in a slice along $\theta$ [see Fig.~\ref{fig:illustration}(c)] is a direct signature of nonadiabatic effects. For the sake of consistency, it can be verified that in the adiabatic limit $\tauB$ converges to the Keldysh time $\sqrt{2 I_p}/F_0$~\cite{Keldysh1965}, and Eq.~\eqref{eq:distribution_z} results in the well-established Ammosov-Delone-Krainov ionization rate~\cite{Ammosov1986, Arissian2010}.

Most importantly, we emphasize that the relative dispersion of the actual $(p,p_z)$-dependent tunneling times for a given barrier configuration is small, below $1\%$ in typical attoclock conditions, see SM. This justifies the characterization of the tunneling dynamics by a single representative \tauB for each $\theta$.

With these observations, we show that the dynamics of the tunneling process {\it in each photoemission direction} $\theta$ can be fully read out from the measurements of the PMD, through Eq.~\eqref{eq:t0} for
the value of \tion and  
by fitting gaussian functions to the  transverse PMDs according to Eq.~\eqref{eq:distribution_z} or Eqs.~\eqref{eq:distribution_p},\eqref{eq:width_dt} for the value of \tauB, as illustrated in Fig.~\ref{fig:illustration}(c).
In terms of accuracy, the corresponding time resolution is of the order of $\delta \tau [{\rm \asec}] {\approx} 0.01 {\times} \delta \theta [{\rm deg}] {\times} \lambda  [{\rm nm}]$, where $\lambda$ is the driving wavelength and $\delta \theta$ is the angular resolution. A typical experimental resolution of $\delta \theta{=}2^{\circ}$~\cite{Eckle2008}  with a standard Ti:Sapphire laser ($\lambda{=}800$~nm) hence corresponds to a temporal resolution $\delta \tau{=}16$~\asec.

To further appraise our reinterpretation of the Attoclock scheme, we applied it to results of numerical simulations and to available experimental data. %
We first confronted the two independent ways of extracting \tauB out of the PMD. Figure~\ref{fig:tunneling_times}(a) shows the correlation between the widths of $\PMD_z(p_z;\theta)$ and $\PMD_p(p;\theta)$, through $\omega\tauB$ and $\omega\tilde\tauB$ respectively, in various theoretical cases. The empty square symbols correspond to results extracted from the PMD obtained in atomic hydrogen, upon numerical resolution of the 3D TDSE, see Methods. Here, we used pulses with a driving wavelength $\lambda=800$ nm, a $\sin^4$ envelop on $\mathbf{A}(t)$ lasting 4 cycles, corresponding to a full duration of 10.68~fs, and an intensity $I_L$ ranging from $0.08$ to $0.2\; \rm PW \ cm^{-2}$, see labels on the r.h.s of Fig~\ref{fig:tunneling_times}(b). Each marker in the plot corresponds to a given orientation in the PMDs, see the $\theta$-scale of  Fig~\ref{fig:tunneling_times}(b).  These data are localised between the diagonal (dashed line), which corresponds to the adiabatic limit ($\omega\tauB=\omega\tilde\tauB$), and the values given by Eq.~\eqref{eq:width_dt} (solid line). The equivalent results obtained in the same conditions but for a short-range potential (circles) are in almost perfect agreement with the solutions of Eq.~\eqref{eq:width_dt}. 

We emphasize that the presence of the Coulomb potential is not a prerequisite for the measurements of \tauB and \tion from the PMDs, in contrast to the original attoclock scheme. However, it clearly affects the values of the extracted times without hampering the validity of our SFA-based interpretations. 
Note that the Coulomb effects do not manifest evenly in the $p_z$ and $p$ directions. The component of the photoelectron velocity is smaller -- and therefore the impact of the Coulomb effect larger -- along $p_z$. Consistently, we observe in Fig.~\ref{fig:tunneling_times}(a) that the numerical data in H are vertically up-shifted compared to the ones obtained with the short-range potential. 
We further verified that the Coulomb potential plays a significantly smaller role for larger $I_p$ or longer driving wavelengths and that the results tend towards the SFA predictions in these limits, see SM. 
A direct practical and general consequence is that the tunneling times extracted from the PMD along $p$ [Eq.~\eqref{eq:distribution_z}] are expected to be more reliable than along  $p_z$  [Eq.~\eqref{eq:distribution_p}].

Therefore, we reported in Fig.~\ref{fig:tunneling_times}(b) the values of \tauB inferred from the fit along $p$, in the H simulations. They are plotted as solid squares, for the four intensities considered in our simulations, against the photoemission angle $\theta$ and the equivalent ionisation time $t_0$, see Eq.~\eqref{eq:t0}. The shaded areas indicate the uncertainties provided by the fitting procedure.  We note that the order of magnitude of \tauB is of few 100 \asec, as predicted by the standard tunneling theories~\cite{Keldysh1965, PerelomovII1967, Buttiker1982, Yudin2001, Eckle2008, Barth2011, Li2017} for the corresponding barriers, see Fig. 2(c) and in agreement with the ones obtained from bichromatic HHG measurements in~\cite{Pedatzur2015}. 
For a given intensity, the tunneling duration increases when the {\it instantaneous} field amplitude decreases, i.e., when $\vert\theta\vert=\omega\vert\tion\vert$ increases. This is because the barrier gets thicker and taller and therefore the duration longer, see Eq. (\ref{eq:Buttiker}). For the same reason, at a given emission angle $\theta$, \tauB increases when \IL decreases. 
The theoretical values given by Eq.~\eqref{eq:gamma_phi} are in excellent agreement with the reconstructed ones. 
Note that the values of $\tauB$ extracted from $\mathcal{D}_z$ are up-shifted by a few tens of asec due to the above-mentioned Coulomb effects, see SM.

\begin{figure}[t]
    \centering
    \includegraphics[width=.4\textwidth]{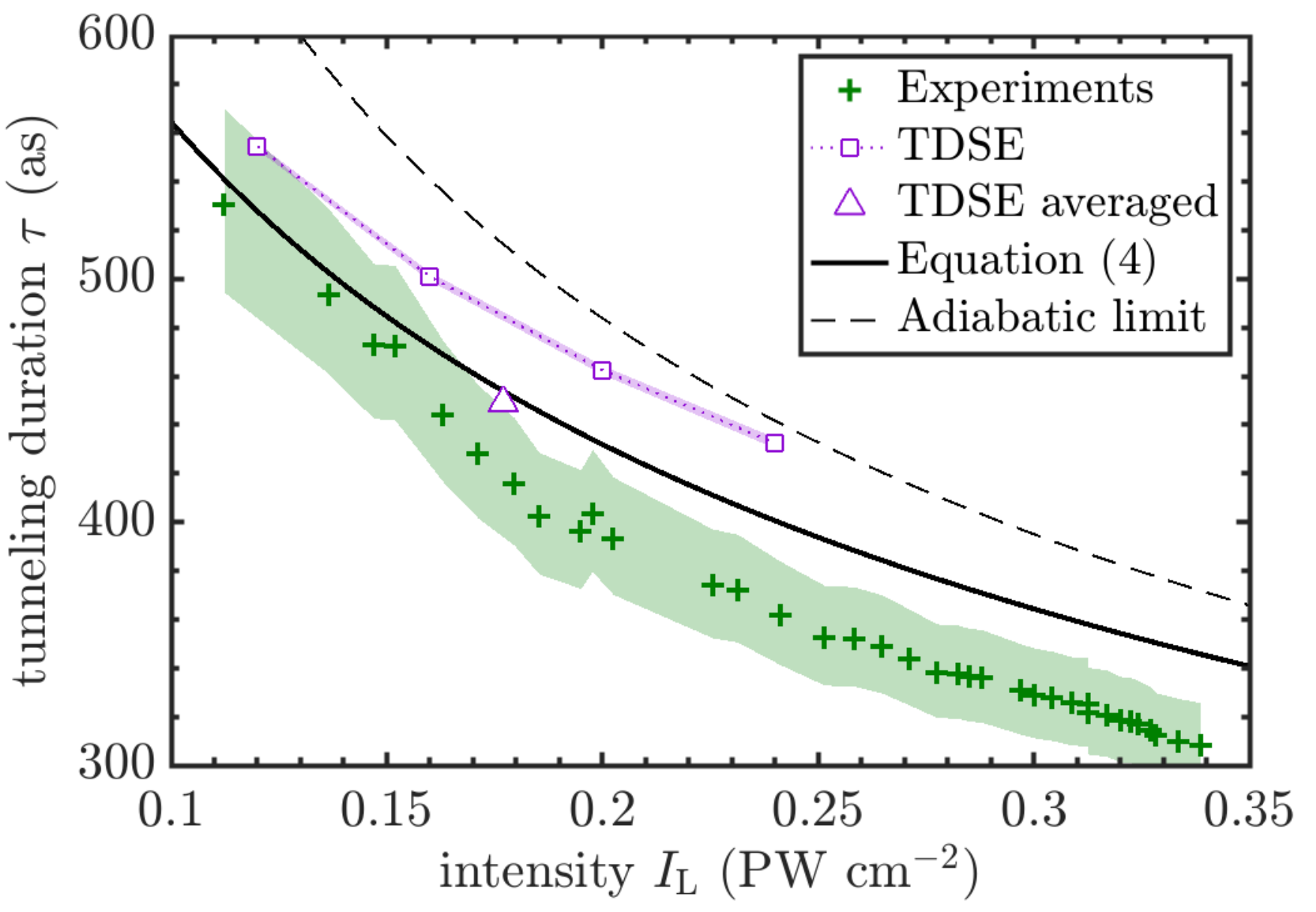}
    \caption{Tunneling durations \tauB (green crosses) in experimental strong field ionization of Ar as a function of the laser intensity. The durations were extracted out of the widths of the PMDs published in Ref.~\cite{Arissian2010} using Eq.~\ref{eq:distribution_p}. The green shaded area corresponds to the published statistical error on the PMD width.
    The experimental data are compared to various theoretical results: numerical solutions of  Eq.~\eqref{eq:gamma_phi} (solid line); adiabatic limit (dashed line); data extracted from the resolution of the TDSE on a model Ar atom (empty squares).  The triangle indicates the value obtained out of the TDSE PMDs averaged over an experimentally realistic intensity  distribution centered at $\IL{=}0.18 \; \rm PW \ cm^{-2}$.} 
    \label{fig:Arissian}
\end{figure}

Eventually, we applied our retrieval procedure to the experimental data published in~\cite{Arissian2010}. In this work, a gaz of Argon (Ar) atoms was ionized by 800-nm circularly polarized laser lasting 15 fs (full width at half maximum) at various intensities. With such long pulses, the obtained PMDs display no anisotropy. At the atomic level, this means that the electrons experienced a single barrier configuration, regardless of the photoemission direction. Moreover, in these experiments only the PMD along $p_z$ is available. Thus, using Eq.~\eqref{eq:distribution_z}, we  extracted a single value of \tauB for each  intensity, displayed as green crosses in Fig.~\ref{fig:width_tunneling_Ar}.  It is of the order of few 100 \asec and decays monotonically as the laser intensity increases, in  good agreement with the ones from Eq.~\eqref{eq:gamma_phi}, displayed as a solid black line and from TDSE (purple squares).  The observed discrepancies can be related to the laser intensity profile within the interaction region which broadens the effectively measured PMDs. To account for this effect, we averaged the TDSE results over a typical \IL distribution and obtained a value (purple triangle) which lies $\sim 20$ \asec closer to the corresponding experimental one. 

To summarize, we have developed a procedure to clock unambiguously the tunneling process inherent to strong field ionization,  employing a circularly polarized light pulse as an attosecond-resolved temporal prism. Using a comprehensive remodeling of the Attoclock, we established that the electron wavepacket emitted {\it in each photoemission direction}, corresponding to a given exit time, carries the value of its own {\it tunneling duration} in the width of its transverse momentum distributions.
This essential result is the direct consequence of the transversal momentum squeezing, a universal property of wavepacket tunneling (in more than one dimension) on which our work sheds light.
When applying our procedure to numerical simulations  and to previously published experimental PMDs, we found finite tunneling durations of few hundred attoseconds in quantitative agreement with the B\"{u}ttiker-Landauer formula and consistent with strong-field theories, thus solving a decade long fundamental puzzle in attosecond science. 

\subsection{Acknowledgments}
We are gratefully thankful to Yann Mairesse and Letizia Fede for valuable inputs and exchanges on the experimental feasibility of tunneling measurements within the Attoclock scheme. We acknowledge Philippe Balcou for helpful discussions on the tunneling time problem in strong-field physics. 
This research received the financial support of the French National Research Agency through Grants No. ANR-20-CE30-0007-DECAP and No. ANR-21-CE30-0036-03-ATTOCOM.
JD acknowledges funding from the European Union’s Horizon Europe research and innovation program under the Marie Sk{\l}odowska-Curie Grant Agreement No. 101154681. Views and opinions expressed are, however, those of the author(s) only and do not necessarily reflect those of the European Union or the European Commission. Neither the European Union nor the European Commission can be held responsible for them.

\subsection{Contributions}
JD derived the main results.
JD and LR performed the numerical simulations.
All the authors discussed and participated in writing the article.
RT, CL, JC  supervised the study and the writing of the article.

\appendix

\section{Methods}
Numerical simulations were performed by solving numerically the TDSE for single-active electron (model) atoms in 3D.  Time-propagation was implemented using an FFT-based split-operator scheme~\cite{Numerical_Recipes} of the fourth order BM4~\cite{McLachlan2022}, with 4096 time-steps on a total duration of 4 laser cycles and a Cartesian position grid. We have used 2048 points from $-400$ a.u. to $400$ a.u. along $x$ and $y$, and 1024 points from $-200$ a.u. to $200$ a.u. along $z$. 
Interaction with the IR field was implemented in the velocity gauge with $\mathbf{A}(t)$ circularly polarized in the $(x,y)$ plane. The latter was assigned a short $\cos^4$ envelop reaching its maximum when aligned along $x$ (i.e. $\theta=0$, see Fig.~\ref{fig:illustration}). 
Orientation-resolved photoelectron spectra were computed using FFT analysis of the continuum wave packets.

We considered three atomic systems. Figures~\ref{fig:tunneling_times}(a) and ~\ref{fig:tunneling_times}(a) show results obtained for the ``exact'' hydrogen atom with Coulombic electron-nucleus interaction, $V(r){=}-1/r$ (in a.u.), where $r$ is the electron-nucleus distance.  Figure \ref{fig:tunneling_times}(a) also shows results with a pseudo-hydrogen based on the short-range potential $V(r) {=} -A\exp ({-} r^2)$. The parameter was set to achieve a ground state energy of $-0.5$ a.u. (corresponding to $\Ip{=}13.6$ eV).
Figure~\ref{fig:width_tunneling_Ar} shows results obtained with an Ar model atom based on a soft-Coulomb potential  $V(\mathbf{r}) = -(r^2+b^2)^{-1/2}$ with the regularisation parameter so that its ground state lies at $-0.58$ a.u. ($\Ip{=}15.7$ eV). All the simulations were performed starting from the ground state. For the Ar case, we verified with SFA-based simulations that the symmetry of the initial state plays a minor role in the obtained results, see also~\cite{Dubois2024}.

\section{Supplementary material}

\subsection{Derivation of the main results}
\subsubsection{PMD in the strong-field approximation}
The objective is to characterize the tunneling dynamics in circularly polarized strong field ionization with a single tunneling time for each photoemission direction $\theta$. The starting point is the strong-field approximation~\cite{Lewenstein1994} which assigns the following ansatz to the continuum electron wave packet
\begin{subequations}
\label{eq:SFA}
\begin{equation}
\label{eq:varphi_LF}
    \varphi (\mathbf{p}) =  \int_{-\infty}^{\infty} \rmd t \; \exp \Big( \rmi S(\mathbf{p},t) \Big) \, s \big( \mathbf{p} + \mathbf{A}(t),t \big) ,
\end{equation}
involving the source term 
\begin{equation}
\label{eq:source_LF}
    s (\mathbf{p},t) = \exp ( \rmi I_p t) \big( \mathbf{r} \cdot \mathbf{F}(t) \big) \, \psi_0 (\mathbf{p}) ,
\end{equation}
and the classical action%, also referred to as the Hamilton principal function~\cite{Goldstein}, 
\begin{equation}
    S (\mathbf{p},t ) = -\dfrac{1}{2} \int_{t}^{\infty} \Big( \mathbf{p} + \mathbf{A}(t^{\prime}) \Big)^2 \rmd t^{\prime} .
\end{equation}
\end{subequations}

The integral in the right-hand side of~\eqref{eq:varphi_LF} is then approximated using the saddle-point approximation~\cite{Lewenstein1994}, 
\begin{equation}
\label{eq:varphi_LF2}
    \varphi (\mathbf{p}) \simeq  \int_{-\infty}^{\infty} \rmd \tion \; C(\mathbf{p},t^\star)\exp \Big( \rmi S(\mathbf{p},t^\star) + \rmi \Ip t^\star \Big) ,
\end{equation}
where 
\begin{equation}
\label{eq:C}
    C(\mathbf{p},t^\star)= \dfrac{|s (\mathbf{p}+\mathbf{A}(t^{\star}),t^{\star})|}{\left| \mathbf{F}(t^{\star}) \cdot (\mathbf{p} + \mathbf{A} (t^{\star})) \right|^{1/2}} ,    
\end{equation}
i.e., the integral is reduced to the dominant contributions at times $t^\star=t_0+i\taup$ solutions of
\begin{equation}
\label{eq:saddle_point}
    \dfrac{1}{2} \Big( \mathbf{p} + \mathbf{A}(t^{\star}) \Big)^2 = - I_p .
\end{equation}
The real part $t_0$ and imaginary part $\taup$ of $t^\star$ were interpreted in the seminal work by Perelomov, Popov and Terent'ev ~\cite{PerelomovII1967} as the ``time of emergence'' from the barrier and the tunneling duration respectively. Equation~\eqref{eq:tstar} in the main text is the transposition of that formalism involving the {\it most probable} value of $\tau^*$, noted \tauB.
By expressing the momentum in cylindrical coordinates,
\begin{subequations}
\begin{equation}
    \mathbf{p} = - p \Big( \mathbf{e}_x \cos \theta + \mathbf{e}_y \sin \theta \Big) + p_z \, \mathbf{e}_z ,
\end{equation}
and introducing its rescaled components
\begin{equation}
    \sigma = \dfrac{\omega}{F(\tion)} p , \qquad \sigma_z  = \dfrac{\omega}{F(\tion)} p_z,
\end{equation}
the slowly varying envelop approximation,  $F(\tion{+}\rmi \taup) {\approx} F(\tion)$, leads to the two conditions
\begin{eqnarray} \label{eqn:condt0}
    && \tion = \theta / \omega , \\
\label{eqn:condtau}
    && 2 \sigma \cosh ( \omega \taup ) = \sigma^2 + \sigma_z^2 + \gamma (\tion)^2 + 1 , \label{eq:phi_sigma}
\end{eqnarray}
\end{subequations} 
for the solutions of Eq.~\eqref{eq:saddle_point}, where
 $\gamma(\tion) = \omega \sqrt{2 I_p} / F(\tion)$ is the instantaneous Keldysh parameter. Note that, while \tion is uniquely defined by the photoemission direction $\theta$ through Eq.~\eqref{eqn:condt0}, \taup is intricately assigned a value for each $\theta$ {\it and each} $(p,p_z)$ through Eq.~\eqref{eqn:condtau}. 

In order to bring forward simple dependencies of the PMD $\PMD (\mathbf{p}) = | \varphi ( \mathbf{p} ) |^2$ on the tunneling durations, we substitute these conditions to the wave packet expressed in Eq.~\eqref{eq:varphi_LF}. We furthermore expand $\varphi( \mathbf{p})$ in each photoemission direction $\theta=\omega\tion$ around its maximum amplitude located at $p=p_0$ and $p_z=0$, up to the second order in $(p-p_0)$ and in $p_z$ respectively. Consistently with the notations in the main text, we call \tauB the value of \taup associated with the dominant  contribution at $(p_0,0)$ for each $\theta$. We then obtain, after several steps of mathematical (and somehow tedious) rewriting, 
a workable expression of the form
\begin{subequations}
\begin{widetext}
\begin{eqnarray*}
    && \exp \Big(   \rmi S (\mathbf{p} , t^{\star} ) + \rmi I_p t^{\star} \Big) \propto 
    \exp \left(  \rmi \dfrac{F(\tion)^2 \tion}{2 \omega^2} \left( \sigma^2 +\gamma^2 + \sigma_z^2 + 1 \right) \right) \nonumber  \\
    &&
    \times  
    \exp \left( - \dfrac{F(\tion)^2}{\omega^2} \taupol (\sigma -\sigma_0)^2 - \dfrac{F(\tion)^2}{\omega^2} \tauB \sigma_z^2  \right) 
     %\qquad\qquad\qquad 
     \times  \exp \Bigg( - \dfrac{2 F (\tion)^2}{\omega^3} \sinh (\omega \tauB) \left( \cosh (\omega \tauB) - \dfrac{\sinh (\omega \tauB)}{\omega \tauB} \right) \Bigg) ,  \label{eqn:phiapprox}
\end{eqnarray*}
\end{widetext}
where $\sigma_0=(\omega/F(t_0))p_0$ is the rescaled most probable momentum
and \taupol is related to \tauB at each $\theta$ following Eq.~\eqref{eq:width_dt}. We can then specialize Eq.~\eqref{eqn:condtau} with the extremum condition to obtain direct mappings of \tauB on $\gamma(\tion)$ on the one hand and on $\sigma_0$ on the other hand,
\begin{eqnarray}
\label{decadix}
     && \gamma (\tion)^2 = \dfrac{\sinh (2 \omega \tauB)}{\omega \tauB} - \dfrac{\big(\sinh ( \omega \tauB ) \big)^2}{(\omega \tauB)^2} - 1  , \\
     && \sigma_0 = \dfrac{\sinh (\omega \tauB)}{\omega \tauB} .
\end{eqnarray}
\end{subequations}
By inserting these expressions in Eq.~\eqref{eqn:phiapprox}, we obtain the main results given in Eq.~\eqref{eq:distribution_z} and Eq.~\eqref{eq:distribution_p} in the main text.

The relative dispersion of the actual $(p,p_z)$-dependent tunneling times for a given $\theta$ 
reads from Eq.~\eqref{eq:phi_sigma} and the definition of \tauB
\begin{equation}
    \dfrac{\Delta \tau}{\tau} \approx \dfrac{\omega^2}{4 F^2 \: \tau \sinh^2 (\omega \tau)}.
\end{equation}
Typically, in ionization from H with a short 800-nm laser, it is below 1\% in the intensity range considered in Fig.~\ref{fig:tunneling_times} (from $0.4\%$ at $0.08 \; \rm PW \ cm^{-2}$ to 0.9\% at $0.2 \; \rm PW \ cm^{-2}$).

\subsubsection{Robustness in the long-pulse limit}
In order to complete the analysis, we consider the long pulse limit, there are ATI peaks and interferences which may change the Attoclock angle~\cite{Hofmann2021}, but it does not change the width, as we demonstrate below.
We consider $F(\tion) {=} F$ constant.
We must take into account the phase of the wavepackets ionized at $\tion$, which interfere at the detector.
The photoelectron momentum distribution is given by
\begin{widetext}
\begin{eqnarray}
    \PMD (p,p_z;\theta) &\propto& \left| \int_{-\infty}^{\infty} {\rm d} \tion  \exp \left( - \dfrac{\taupol}{2} (p - p_0)^2 - \dfrac{\tauB}{2} p_z^2  \right) \; \delta (\theta - \omega \tion) \; C(\mathbf{p},t^{\star})  \; \exp \left(  \rmi \dfrac{F^2 \tion}{2 \omega^2} \left( \sigma^2 + \sigma_z^2  + 1+\gamma^2 \right) \right) \right|^2 , \nonumber \\
    &\propto&  \int_{- \pi/\omega}^{\pi/\omega} {\rm d} \tion \; \exp \Big( - \taupol (p - p_0)^2 - \tauB p_z^2    \Big)  \; \delta (\theta - \omega \tion) \; |C(\mathbf{p},t^{\star})|^2 \nonumber \\
    && \qquad\qquad\qquad\qquad\qquad\qquad\qquad\qquad\qquad \times \sum_{n{=}-\infty}^{\infty}  \delta \left( n \omega -  \dfrac{F^2}{2 \omega^2} \left( \sigma^2 + \sigma_z^2 + 1+\gamma^2 \right) \right)  .
\end{eqnarray}
\end{widetext}
The resulting photoelectron momentum distribution along $\theta {=} \omega \tion$ is composed of ATI peaks at 
\begin{equation}
    \dfrac{\mathbf{p}_n^2}{2} = n \omega - \dfrac{F^2}{2 \omega^2} \left(1+\gamma^2 \right) ,
\end{equation}
which are modulated by the distribution~\eqref{eq:distribution_p}. The ATI peaks result from the interference between the wavepacket ionized at different laser cycles. For instance, they correspond to the peaks along $\theta {=}180^{\circ}$ in the photoelectron momentum distribution of Fig.~\ref{fig:illustration}. Therefore, our procedure is still applicable, since the fit can still be made to retrieve the tunneling time from $\taupol$ and \tauB, even in the long pulse limit, and even if interference patterns are visible.

\subsection{B\"uttiker-Landauer traversal time and transverse squeezing}

Below, we provide a simple derivation for the description of the momentum squeezing for a 3D particle moving with a total energy $E$ through an arbitrary static barrier $V(x)$ in the $x$ direction (starting from $x=-\infty$), and freely in the transverse plane with momentum  $\mathbf{p}_{\perp}$.
To this end, we consider the time independent Schr\"odinger equation in the $(x,\mathbf{p}_{\perp})$ representation, 
\begin{equation}
     \left( - \dfrac{\hbar^2 \partial_x^2}{2 m} + \dfrac{\mathbf{p}_{\perp}^2}{2 m} + V(x) - E \right) \phi (x,\mathbf{p}_{\perp}) = 0 . 
\end{equation}
The WKB approximation of its solution reads 
\begin{equation}
    \phi (x,\mathbf{p}_{\perp}) \propto \exp \left( \dfrac{\rmi}{\hbar} \int_{-\infty}^x \sqrt{2 m [ E - V(s) ] - \mathbf{p}_{\perp}^2} \; \rmd s\right) .
\end{equation}
Then, considering that the particle's motion is mostly directed along $x$, we expand the solution around $\mathbf{p}_{\perp} {=}\boldsymbol{0}$ and obtain the expression
\begin{eqnarray}
    && \phi (x,\mathbf{p}_{\perp}) \propto \exp \left( \dfrac{\rmi}{\hbar} \int_{-\infty}^x \sqrt{2 m [E - V(s)]} \; \rmd s \right) \nonumber \\
    && \qquad \qquad \times \exp \left( - \dfrac{\rmi}{\hbar}\int_{-\infty}^x \dfrac{\mathbf{p}_{\perp}^2 \; \rmd s}{2\sqrt{ 2 m [E - V(s)]}} \right) . \label{eqn:phiWKB2}
\end{eqnarray}
where the integrands are complex-valued only in the tunneling region $x{\in}[x_{\rm in},x_{0}]$, i.e. where $E {-} V(x) {\leq} 0$. Therefore, the density associated with the wavefunction after tunneling (large $x$) reads explicitly as
\begin{eqnarray}
    && \lim_{x {\to} \infty} | \phi (x,\mathbf{p}_{\perp}) |^2 \propto \exp \left( - \dfrac{2}{\hbar} \int_{x_{\rm i}}^{x_{\rm e}} \sqrt{2 m [ V(s) - E ]} \; \rmd s \right) \nonumber \\
    && \qquad \qquad \times \exp \left( - \dfrac{1}{\hbar} \int_{x_{\rm i}}^{x_{\rm e} } \dfrac{\mathbf{p}_{\perp}^2 \; \rmd s}{\sqrt{ 2 m [ V(s) - E]}} \right) . \label{eqn:asymdenWKB}
\end{eqnarray}
Note that outside the tunneling region, i.e. where $s{\notin}[x_{\rm in},x_{0}]$, the integrals showing up in Eq.~\eqref{eqn:phiWKB2} cancel out by complex conjugation when computing the asymptotic density. 
In Eq.~\eqref{eqn:asymdenWKB}, the argument of the first exponential corresponds to the 1D WKB ionization rate. The argument of the second exponential corresponds to the momentum squeezing. 
More explicitly, if we keep explicit the second term only, we obtain
\begin{equation}
    \lim_{x {\to} \infty} | \phi (x,\mathbf{p}_{\perp}) |^2 \propto  \exp \left( - \dfrac{ \; \mathbf{p}_{\perp}^2}{m \; \hbar} \tauB\right) ,
\end{equation}
where $\tauB$ is the B\"{u}ttiker-Landauer transversal time given in Eq.~\eqref{eq:Buttiker} in the main text. This result shows that transversal squeezing is a general property of quantum systems undergoing tunnel ionization. It also confirms our findings and our description of tunneling time measurements by means of the transversal squeezing phenomenon.

\subsection{Imprint of the Coulomb potential in the tunneling times}

\begin{figure}
    \centering
    \includegraphics[width=.4\textwidth]{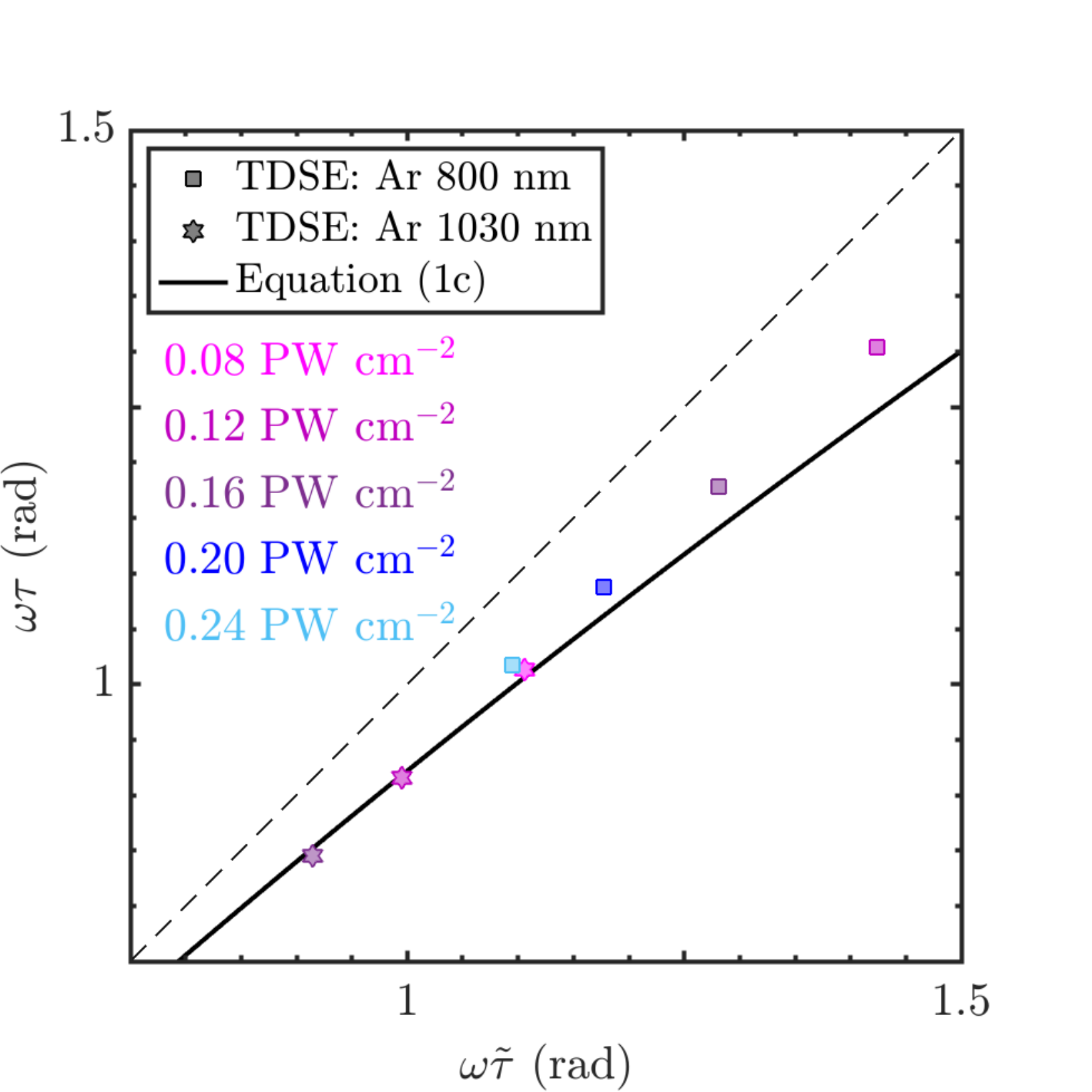}
    \caption{Width of the photoelectron momentum distribution for different $\theta$ modeled by Eq.~\eqref{eq:distribution_p} as a function of the tunneling phase $\omega \tauB$ and the associated tunneling time in \asec for laser wavelengths 800 nm (circles) and 1030 nm (diamonds) of two-laser cycles duration. The markers are the results from three-dimensional simulations with a long-range potential for Ar.}\label{fig:width_tunneling_Ar}
\end{figure}
\begin{figure}
    \centering
    \includegraphics[width=.4\textwidth]{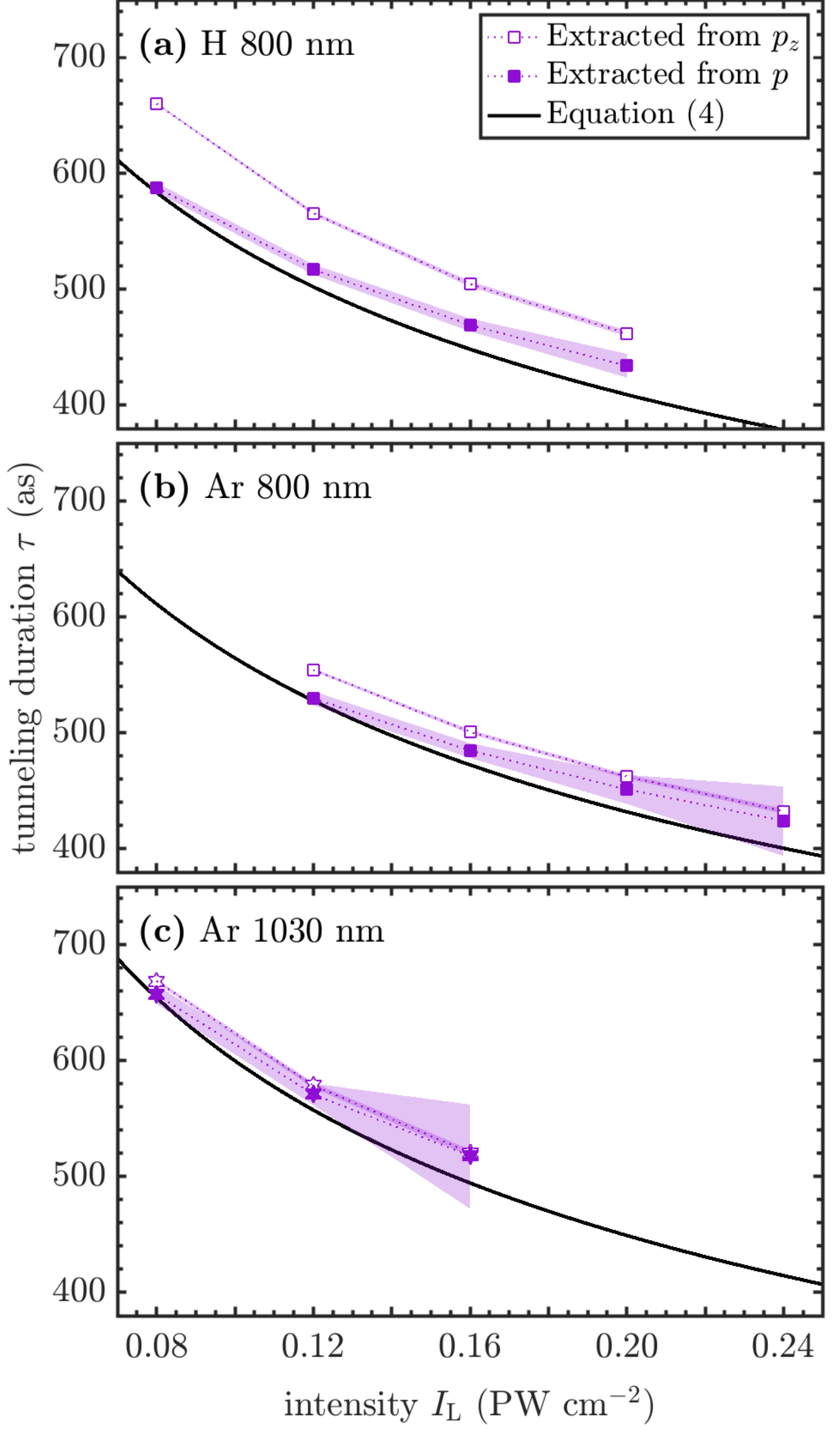}
    \caption{Comparison of the tunneling times in Ref.~\cite{Barth2011} and the ones extracted from the width of the distribution along the propagation axis (red symbols) and in the polarization plane (green symbols). (a) H for 800 nm with four-laser cycles. (b) Ar for 800 nm with two-laser cycles. (c) Ar for 1030 nm with two-laser cycles.}
    \label{fig:tunneling_times_intensity_H_Ar}
\end{figure}

In Figs.~\ref{fig:width_tunneling_Ar} and~\ref{fig:tunneling_times_intensity_H_Ar}, we tested the role of the laser wavelength, either 800 or 1030 nm, on the tunneling duration extracted from the PMD compared to the analytic expression of Eq.~\eqref{eq:width_dt} and found a better agreement as the Keldysh parameter is smaller for longer wavelength at the same intensity. 

\begin{figure*}[t]
    \centering
    \includegraphics[width=.8\textwidth]{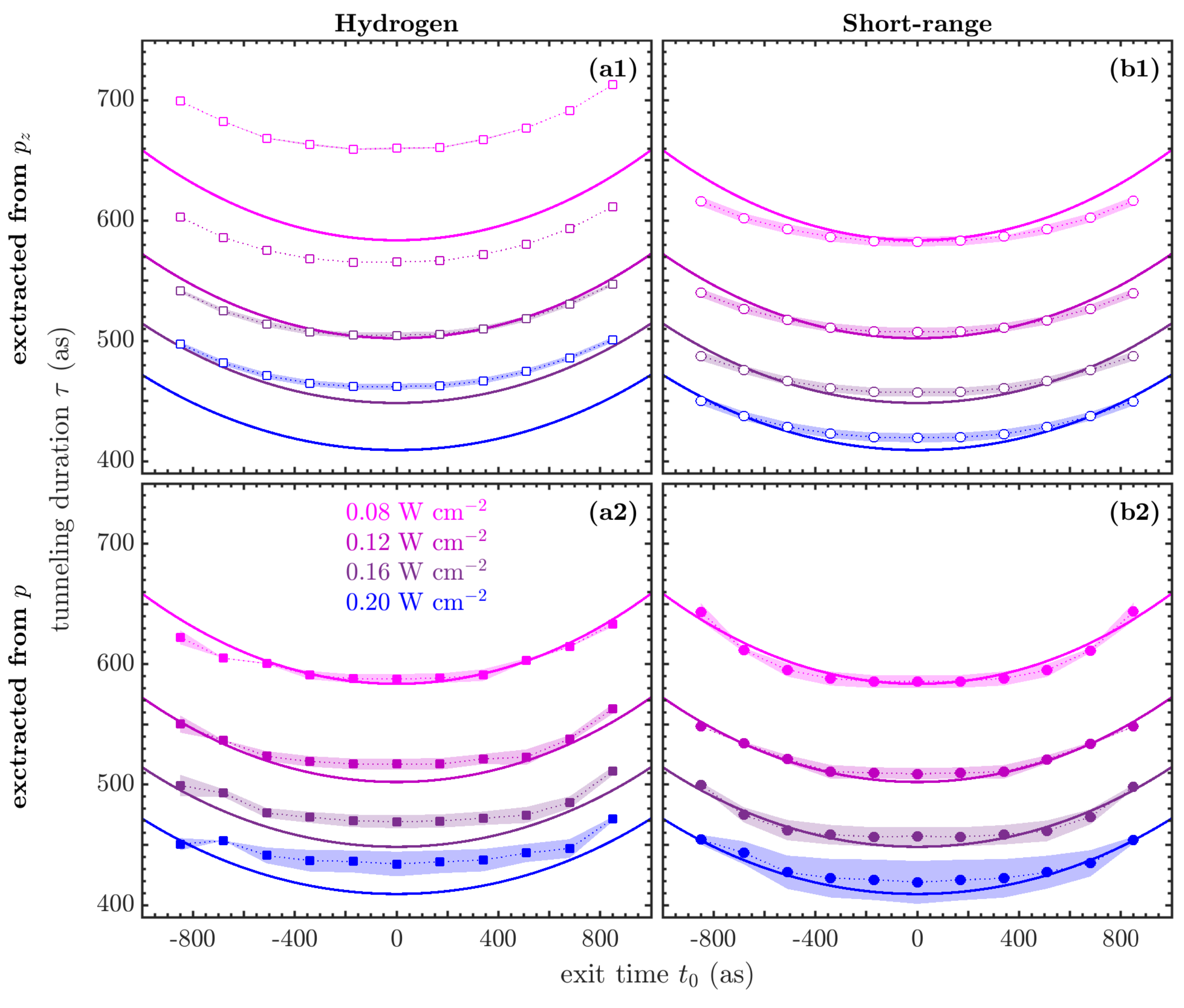}
    \caption{Tunneling time as a function of the angle at 800 nm for different laser intensities obtained from three-dimensional (3D) simulations. The solid lines are the theoretical values~\cite{Barth2011} of the tunneling time (same lines in each panels). The lines with symbols are the tunneling times extracted from the photoelectron momentum distributions using (1) the width along the propagation axis and (2) the width along the instantaneous vector potential. The colored regions are the statistical errors on the fitting parameters. The photoelectron momentum distributions are computed with a (a) Hydrogen potential $-1/|\mathbf{r}|$ and (b) Gausian potential.}
    \label{fig:tunneling_times_gaussian_H}
\end{figure*}

We also tested the case of short- and long-range H with laser wavelength 800 nm. it is displayed in Fig.~\ref{fig:tunneling_times_gaussian_H}.
As could be expected, the most pronounced discrepancies appear at the largest intensity near $\theta{=}0^\circ$. Again, this is where the instantaneous field at ionization reaches its largest values. Then, the essential states used to model the process become significantly affected by dressing effects. Typically, the Stark shift is not accounted for in the derivations of Eqs.~\eqref{eq:distribution_z}-\eqref{eq:width_dt}, which rely on the plain SFA. Including them would require a finer modeling and the input of the accurate value of \IL~\cite{Labeye2018}. Nevertheless, the general agreement between the numerical and theoretical results conforts the validity of our approach. In addition, in Fig.~\ref{fig:tunneling_times_gaussian_H}(a), we observe an asymmetry in the extracted $\tauB$ with respect to $\theta=0$ at large angles. This is a clear signature of the Coulomb asymmetry, that would modify Eq.~\eqref{eq:t0}, since it is absent in the short-range case, see Fig.~\ref{fig:tunneling_times_gaussian_H}(b).

\end{document}